# Intertwined charge and pair density orders in a monolayer high-$T_c$ iron-based superconductor

Tianheng Wei[a,1], Yanzhao Liu[a,1], Wei Ren[a,1], Ziqiang Wang[b] & Jian Wang[a,c,d,2]

[a]*International Center for Quantum Materials, School of Physics, Peking University, Beijing 100871, China*
[b]*Department of Physics, Boston College, Chestnut Hill, MA 02467, USA*
[c]*Collaborative Innovation Center of Quantum Matter, Beijing 100871, China*
[d]*Hefei National Laboratory, Hefei 230088, China*

**Author contributions:** J.W. conceived and supervised the research. Y.L., T.W. and W.R. grew the samples and carried out the STM/STS experiments. T.W., Y.L. and J.W. analyzed the experimental data. T.W., Y.L., Z.W. and J.W. wrote the manuscript.

The authors declare no conflict of interest.

[1]T.W., Y.L. and W.R. contributed equally to this work.

[2]To whom correspondence may be addressed. Email: jianwangphysics@pku.edu.cn (J.W.)

**Abstract**
Symmetry-breaking electronic phase in unconventional high-temperature (high-$T_c$) superconductors is a fascinating issue in condensed-matter physics, among which the most attractive phases are charge density wave (CDW) phase with four unit-cell periodicity in cuprates and nematic phase breaking the $C_4$ rotational symmetry in iron-based superconductors (FeSCs). Recently, pair density wave (PDW), an exotic superconducting phase with non-zero momentum Cooper pairs, has been observed in high-$T_c$ cuprates and the monolayer FeSC. However, the interplay between the CDW, PDW and nematic phase remains to be explored. Here, using scanning tunneling microscopy/spectroscopy, we detected commensurate CDW and CDW-induced PDW orders with the same period of $\lambda = 4a_{Fe}$ ($a_{Fe}$ is the distance between neighboring Fe atoms) in a monolayer high-$T_c$ Fe(Te,Se) film grown on SrTiO$_3$(001) substrate. Further analyses demonstrate the observed CDW is a smectic order, which breaks both translation and $C_4$ rotational symmetry. Moreover, the smecticity of the CDW order is strongest near the superconducting gap but weakens near defects and in an applied magnetic field, indicating the interplay between the smectic CDW and PDW orders. Our works provide a new platform to study the intertwined orders and their interactions in high-$T_c$ superconductors.

**Introduction**

Exploring complex symmetry-breaking electronic orders emerging in high-temperature (high-$T_c$) superconductors and their interplays is a fundamental issue for understanding the correlated high-$T_c$ superconductivity[1–3]. Among the multiple distinct electronic phases, the charge density wave (CDW) and nematic phases have attracted particular attention, which are widely observed in cuprates[4,5] and iron-based superconductors (FeSCs)[2,6,7], respectively. Recently, the pair density wave (PDW) order with a spatially modulating superconducting order parameter has been reported in high-$T_c$ cuprates[8–12] and the monolayer FeSC[13]. Furthermore, the PDW order induced by CDW has been detected in superconducting transition-metal dichalcogenides[14], where the PDW order shows the same wavevector as the CDW order. However, previous studies of PDW mainly focused on detection of the PDW order and exploring the underlying mechanism[8–11,13,15–19]. The interplay among the symmetry-breaking CDW, PDW and nematic phase is rarely discussed.

In the presence of uniform superconductivity, the lowest coupling term between the charge and pair density wave carrying the same momenta is[20]:

$$f_c = \kappa \sum_{j=x,y} \left( \rho_{\mathbf{Q}_j} \Delta_0^* \Delta_{-\mathbf{Q}_j} + \rho_{\mathbf{Q}_j}^* \Delta_0^* \Delta_{\mathbf{Q}_j} \right) + \text{c. c.}$$

where $\Delta_0 = |\Delta_0| e^{i\theta}$, $\rho_{\mathbf{Q}_j} = |\rho_{\mathbf{Q}_j}| e^{i\varphi_{\mathbf{Q}_j}^{CDW}}$, $\Delta_{\pm \mathbf{Q}_j} = |\Delta_{\mathbf{Q}_j}| e^{i\varphi_{\pm \mathbf{Q}_j}^{PDW}}$ are order parameters of uniform superconductivity, CDW and PDW, respectively, $\kappa$ is a coupling coefficient. The interplays between the charge and pair density wave orders are generated by this coupling. Experimentally, the charge density and pair density modulation can be directly visualized by detecting the local density of states (LDOS) and superconducting coherence peak height with scanning tunneling microscope (STM) and scanning tunneling spectroscopy (STS) [9,13]. Here, using STM and STS, we reported a commensurate CDW and the CDW-induced PDW with the same period of $4a_{Fe}$ in a monolayer Fe(Te,Se) film grown on SrTiO$_3$(001) (STO) substrate. The CDW order is found to simultaneously break rotational symmetry, which is carefully analyzed. A rotational symmetry breaking, smectic PDW origin of the $C_4$ symmetry breaking CDW is revealed based on the energy and magnetic field dependent measurements.

**Results**

**$4a_{Fe}$ charge density wave in the monolayer Fe(Te,Se) film.** The one-unit-cell (1-UC) Fe(Te,Se) film was epitaxially grown on a pretreated Nb-doped SrTiO$_3$(001) (STO) substrate (*SI Appendix*). Fig. 1a shows a typical large-scale topography $T(\mathbf{r})$ of the film, which reflects the high-quality of the film by the atomically-flat surface. At terraces inherited from STO substrate, 2$^{nd}$-UC Fe(Te,Se) is grown in step-flow mode. The large widths of terraces (about 300 nm) correspond to a small miscut angle. As shown in the inset of Fig. 1a, the thickness of the 2$^{nd}$ layer is around 0.58 nm, corresponding to the nominal stoichiometry x ≈ 0.7[21] of the FeTe$_{1-x}$Se$_x$ film. An averaged tunneling d$I$/d$V$ spectrum over 67 points along a 33.3 nm linecut is shown in Fig. 1b (waterfall of these spectra is shown in Fig. S1). The spectra are measured at 4.6 K, which is much lower than the superconducting transition temperature[21–23]. The fully gapped feature and two pairs of coherence peaks ($\Delta_1 \approx 11$ meV and $\Delta_2 \approx 16$ meV) are consistent with previous studies[22–24]. Strikingly, as the set-up bias is decreased to $V_s = 40$ mV, an ordered pattern emerges in the topographic image, the directions of the pattern are rotated by 45° to the surface Te/Se lattice, that is, along two Fe-Fe lattice directions (Fig. 1c). This modulated pattern breaks the lattice translational symmetry with a supercell sketched in the inset of Fig. 1d. The magnitude of the Fourier transform of the topography $T(\mathbf{q})$ (Fig. 1d) reveals four peaks at $\mathbf{Q} = (\pm 0.25, 0)Q_{Fe}$ and $(0, \pm 0.25)Q_{Fe}$ ($Q_{Fe} = 2\pi/a_{Fe}$), corresponding to a modulation with a period of $4a_{Fe}$ along the two Fe-Fe bond directions. The broadening of peaks can be attributed to the randomly distributed ordered domains.

To investigate the energy dependent electronic properties of the $4a_{Fe}$ modulation, d$I$/d$V$ maps at different bias voltages $g(\mathbf{r}, V) \equiv dI/dV(\mathbf{r}, V)$ are measured at the same area in Fig. 1c. All d$I$/d$V$ maps are drift-corrected according to the simultaneously obtained topography by the Lawler-Fujita method[25]. Fig. 1e shows the d$I$/d$V$ map at 10 mV, which is close to the superconducting coherence peak. The $4a_{Fe}$ modulation is clearly observed in real space map $g(\mathbf{r}, 10\ mV)$ and further confirmed by its Fourier transform $g(\mathbf{q}, 10\ mV)$ (Fig. 1f), which shows Fourier peaks at $\mathbf{Q}_{x,y} = (\pm 0.25, 0)Q_{Fe}$ and $(0, \pm 0.25)Q_{Fe}$, respectively. Linecut of Fourier transforms along $x$ direction (red dashed line in Fig. 1f) at different bias voltages is shown in Fig. 1g. The modulation wavevector does not disperse with energy, demonstrating the charge order nature of the modulation rather than quasiparticle interference (QPI)[23]. The modulation is invisible at Fermi energy and emerges when non-zero LDOS is induced by increasing the bias voltage. Intriguingly, the modulation signal is maximized around the superconducting coherence peak and becomes weaker as the bias voltage increases beyond the superconducting gap energy (more information can be found in Fig. S2).

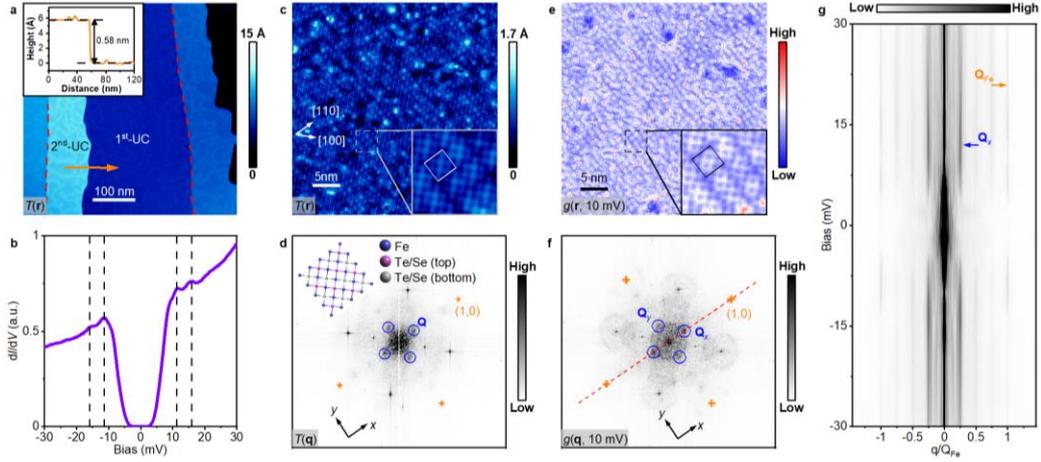

**Fig. 1. $4a_{Fe}$ charge density wave order in monolayer Fe(Te,Se)/STO. a**, A topographic image of the Fe(Te,Se)/STO film (500×500 nm$^2$, $V_s$ = 3 V, $I_s$ = 80 pA). Two terraces inherited from STO substrate are marked by red dashed lines. 2$^{nd}$-UC Fe(Te,Se) grows at terraces in step-flow mode. The inset shows the thickness of the 2$^{nd}$-UC Fe(Te,Se) measured along the orange arrow. **b**, An averaged tunneling d$I$/d$V$ spectrum over 67 points along a 33.3 nm linecut ($V_s$ = 40 mV, $I_s$ = 500 pA, $V_{mod}$ = 0.8 mV) (Fig. S1). Two pairs of dashed lines denote the superconducting coherence peaks ($\Delta_1 \approx 11$ meV and $\Delta_2 \approx 16$ meV). **c**, A topographic image with atomic resolution (37.1×38.6 nm$^2$, $V_s$ = 40 mV, $I_s$ = 100 pA). The inset is the zoom-in topography clearly showing the charge density modulations along two Fe-Fe directions, with a supercell marked by white rectangle. **d**, The magnitude of the Fourier transform of **c**. Bragg peaks of Fe lattice are denoted by orange crosses. The CDW wavevectors $\mathbf{Q} = (\pm 0.25, 0)Q_{Fe}$ and (0, ±0.25)$Q_{Fe}$ are circled in blue. Inset: the schematic of one $4a_{Fe} \times 4a_{Fe}$ supercell corresponding to the charge order. **e**, d$I$/d$V$ map at 10 mV at the same area of **c** ($V_s$ = -100 mV, $I_s$ = 2 nA, $V_{mod}$ = 0.6 mV). Inset: the zoom-in d$I$/d$V$ map showing the LDOS modulation with a supercell marked by black rectangle. **f**, The magnitude of Fourier transform of **e**. Orange crosses are at (±$Q_{Fe}$, 0) and (0, ±$Q_{Fe}$). CDW wavevectors at $Q_{x,y}$ = (±0.25, 0)$Q_{Fe}$ and (0, ±0.25)$Q_{Fe}$ in $x$ and $y$ directions are circled in blue. **g**, Linecut of Fourier transforms along $x$ direction (red dashed line in **f**) at different bias voltages. Non-dispersive CDW wavevector $Q_x$ = 0.25 $Q_{Fe}$ is marked by blue arrow.

**Anisotropy of the $4a_{Fe}$ CDW.** Different from the underdoped cuprates that usually show electronic checkerboard patterns[4,5], most FeSCs exhibit nematic order[2,6,7], which breaks the $C_4$ rotational symmetry, such as the stripe charge orders in non-superconducting multi-layer FeSe/STO[26]. Due to the existence of quenched disorder, the observed CDW symmetry in our monolayer Fe(Te,Se)/STO is hard to discriminate between the stripe and checkerboard order directly from the real-space d$I$/d$V$ map[27]. Instead, the Fourier transform results provide more information on the anisotropy of the observed CDW order. It is noticed that in Fig. 1f, Fourier peaks of the $4a_{Fe}$ CDW in $x$ direction $Q_x$ are stronger than that in $y$ direction $Q_y$ (Fig. S3), indicating the $C_4$ symmetry is broken. The 2D lock-in analysis[10,11,13,28] is applied to reveal the spatial variation of modulations along two directions. The local complex amplitude of the modulation at $Q_{j=x,y}$ can be estimated by $A^C_{Q_j}(\mathbf{r}) = \int d\mathbf{R} C(\mathbf{R}) e^{i\mathbf{Q}_j \cdot \mathbf{R}} e^{-\frac{(\mathbf{r}-\mathbf{R})^2}{2\lambda^2}}$, where $C(\mathbf{R})$ is an arbitrary real space image and $\lambda = 2\pi/Q$ is the period of the modulation. The local modulation amplitude $|A^C_{Q_j}(\mathbf{r})|$ and phase $\phi^C_j(\mathbf{r})$ are obtained by $|A^C_{Q_j}(\mathbf{r})| = \sqrt{\text{Re } A^C_{Q_j}(\mathbf{r})^2 + \text{Im } A^C_{Q_j}(\mathbf{r})^2}$ and $\phi^C_{Q_j}(\mathbf{r}) = \text{Arg}(A^C_{Q_j}(\mathbf{r}))$, respectively. Figs. 2a-b plot the modulation amplitudes $|A^g_{Q_{x(y)}}(\mathbf{r}, 10 \text{ mV})|$ of $g(\mathbf{r}, 10 \text{ mV})$ at $Q_x$ and $Q_y$ with the same colorbar scale. Randomly distributed modulation domains can be clearly distinguished by red patches in the amplitude maps. Note that the modulation amplitude in $x$ direction is stronger than $y$ direction in most part of the map. To characterize the local unidirectionality of the modulation, a directional order parameter map[29] $N^g(\mathbf{r}) = \frac{|A^g_{Q_x}(\mathbf{r})| - |A^g_{Q_y}(\mathbf{r})|}{|A^g_{Q_x}(\mathbf{r})| + |A^g_{Q_y}(\mathbf{r})|}$ is shown in Fig. 2c. The modulation in $x$ direction dominates in almost all the measured area (red area where $N^g(\mathbf{r}) > 0$), while only small regions are dominated by $y$ direction modulation (blue area where $N^g(\mathbf{r}) < 0$). The strong anisotropy indicates a $C_4$ symmetry breaking stripe order rather than a checkerboard order. Figure 2d shows the energy dependence of the normalized modulation intensity along $x$ and $y$ directions, where the normalized intensity is defined as the integrated intensity around Fourier peaks $Q_{j=x,y}$ in $g(\mathbf{q}, V)$ divided by the integrated intensity of the whole $g(\mathbf{q}, V)$. The energy-dependent anisotropy, i.e., difference between the modulation intensity along two directions shows a peak near the superconducting gap energy, indicating a potential relationship between the anisotropy and the superconductivity.

Apart from the comparison of the modulation intensity along $x$ and $y$ directions, more quantitative criteria have been proposed to distinguish stripe and checkerboard order under circumstances in which quenched disorders obscure the difference[27]. As shown in Figs. 2e-f, the energy dependence of three quantities $\xi_{CDW}$, $\xi_{orient}$ and $\eta_{orient}$ are calculated, which are defined in terms of the complex modulation amplitudes $A^g_{\mathbf{Q}_j}(\mathbf{r})$ mentioned above:

$$\xi^2_{CDW} \equiv \frac{\left|\int d\mathbf{r} A^g_{\mathbf{Q}_x}\right|^2 + \left|\int d\mathbf{r} A^g_{\mathbf{Q}_y}\right|^2}{\int d\mathbf{r} \left(\left|A^g_{\mathbf{Q}_x}\right|^2 + \left|A^g_{\mathbf{Q}_y}\right|^2\right)},$$

$$\xi^2_{orient} \equiv \frac{\left|\int d\mathbf{r} \left(\left|A^g_{\mathbf{Q}_x}\right|^2 - \left|A^g_{\mathbf{Q}_y}\right|^2\right)\right|^2}{\int d\mathbf{r} \left(\left|A^g_{\mathbf{Q}_x}\right|^2 - \left|A^g_{\mathbf{Q}_y}\right|^2\right)^2},$$

$$\eta_{orient} \equiv \frac{\int d\mathbf{r} \left(\left|A^g_{\mathbf{Q}_x}\right|^2 - \left|A^g_{\mathbf{Q}_y}\right|^2\right)^2}{\int d\mathbf{r} \left(\left|A^g_{\mathbf{Q}_x}\right|^2 + \left|A^g_{\mathbf{Q}_y}\right|^2\right)^2}.$$

$\xi_{CDW}$ and $\xi_{orient}$ plotted in Fig. 2e have units of length, characterizing the correlation lengths of the $4a_{Fe}$ CDW and its directionality, respectively. $\eta_{orient}$ is dimensionless and characterizes the stripiness. Theoretically, when $\xi_{CDW}$ is smaller than $4\lambda$ (grey region in Fig. 2e), a fluctuating order dominates and it is hard to distinguish between stripe and checkerboard orders in this energy range. At around the superconducting gap energy, $\xi_{CDW}$ is larger than $4\lambda$ and a static order is expected. Furthermore, around the gap energy, the directional correlation length $\xi_{orient}$ is much longer than $\xi_{CDW}$ and the stripiness $\eta_{orient}$ (Fig. 2f) is larger than a critical value (0.2), which agrees well with the theoretically predicted features of stripe order[27]. Beyond the superconducting gap energy, $\eta_{orient}$ becomes smaller than 0.2 and $\xi_{orient}$ quickly decays, indicating the restoration of the $C_4$ symmetry, consistent with the fading anisotropy in Fig. 2d.

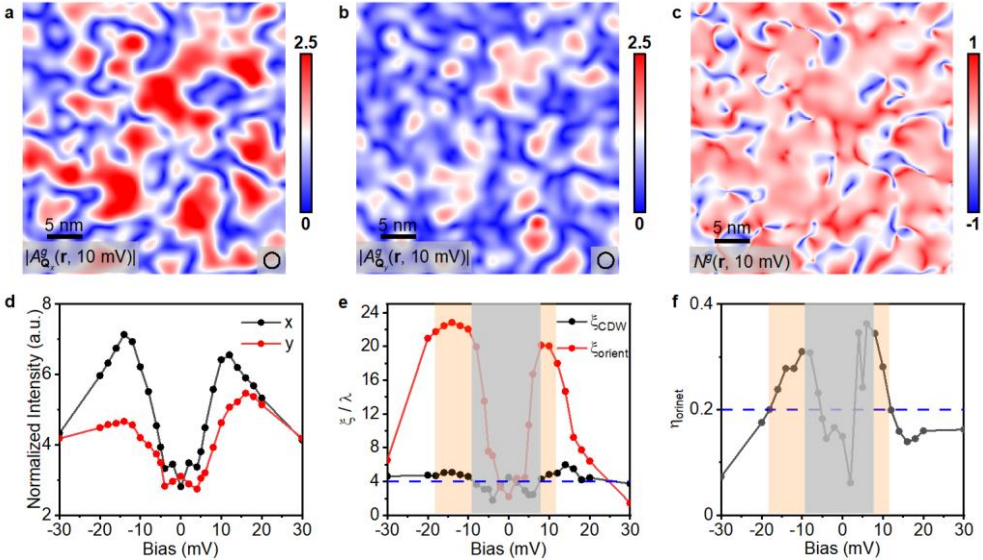

**Fig. 2. Energy dependent $C_4$ symmetry breaking of the CDW order. a,b**, The spatial distribution of the modulation amplitude $\left|A^g_{\mathbf{Q}_{x(y)}}(\mathbf{r}, 10\text{ mV})\right|$ in $x$ and $y$ directions calculated by 2D lock-in method in Fig. 1e, respectively. The colorbar scales in **a** and **b** are the same, with red for strong modulation and blue for weak modulation. The modulation amplitude in $x$ direction is obviously stronger than $y$

direction in most of the area. The black circles denote the averaging length scales with their radius equal to the modulation period $4a_{Fe}$. **c**, The spatial distribution of the directional order parameter

$$N^g(\mathbf{r}, 10\text{ mV}) = \frac{\left|A^g_{\mathbf{Q}_x}(\mathbf{r}, 10\text{ mV})\right| - \left|A^g_{\mathbf{Q}_y}(\mathbf{r}, 10\text{ mV})\right|}{\left|A^g_{\mathbf{Q}_x}(\mathbf{r}, 10\text{ mV})\right| + \left|A^g_{\mathbf{Q}_y}(\mathbf{r}, 10\text{ mV})\right|}.$$ **d**, The energy dependence of the normalized intensity of

CDW modulation peaks in $x$ and $y$ directions. The difference of the two directions peaks near the superconducting gap energy and disappears at 30 mV. **e**, The energy dependence of correlation lengths of the CDW order ($\xi_{CDW}$) and its directionality ($\xi_{orient}$). **f**, The energy dependence of the stripiness $\eta_{orient}$ of the CDW order. Grey areas in **e** and **f** denote energy range where $\xi_{CDW}<4\lambda$ and the directionality is hard to determine. Orange areas in **e** and **f** represent energy ranges where $\xi_{orient}>\xi_{CDW}>4\lambda$ and $\eta_{orient}>0.2$. $C_4$ symmetry breaking stripe order is preferred in this region.

**Induced smectic pair density wave.** As shown in Fig. 2, both the intensity and anisotropy of the $4a_{Fe}$ modulation are strongest near the superconducting gap energy, which may lead to a rotational symmetry breaking (directional) PDW induced by the interaction between the unidirectional CDW order and superconductivity[20]. It has been evidenced that the coherence peak height (CPH) has a strong positive correlation with the superconducting order parameter[9,13,30,31]. Thus, we measure d$I$/d$V$ maps with denser bias voltages (step = 0.5 mV) and extract the CPH at every pixel in the area shown in Fig. 3 (*SI Appendix*). Clear $4a_{Fe}$ modulations of the CPH can be observed in both real space map $H(\mathbf{r})$ (Fig. 3a) and the Fourier transform $H(\mathbf{q})$ (Fig. 3b), which is consistent with a CDW-induced PDW order. Similar to the d$I$/d$V$ modulation, the CPH modulation also shows anisotropy with stronger Fourier peaks in $x$ direction. Figures. 3c-d illustrate tunneling spectra measured along the two linecuts indicated by the black ($x$ direction) and red ($y$ direction) arrows in Fig. 3a. The CPH modulations with a period of $4a_{Fe}$ along two directions as well as the anisotropy (Figs. 3b and 3e) demonstrate the existence of the induced rotational symmetry breaking PDW.

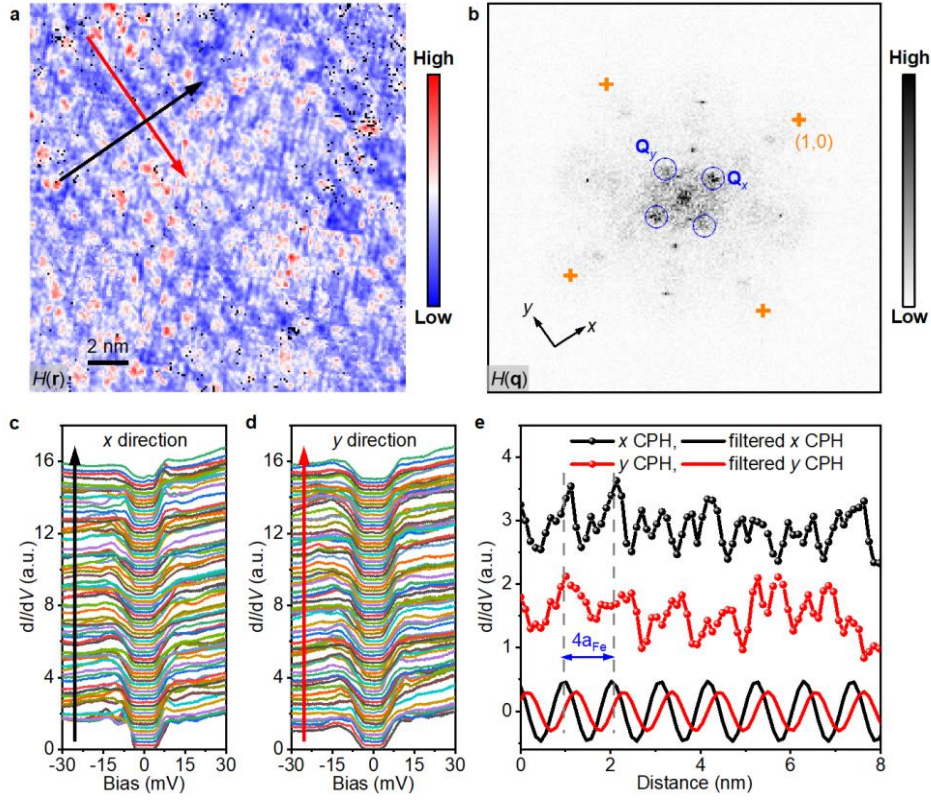

**Fig. 3. Periodic modulation of the coherence peak height. a**, The coherence peak height map $H(\mathbf{r})$ measured over a 18.3×19.1 nm² field of view ($V_s$ = -100 mV, $I_s$ = 6 nA, $V_{mod}$ = 0.6 mV). The coherence peak height shows similar modulation with the CDW order in d$I$/d$V$ maps. **b**, The magnitude of Fourier transform of **a**. Orange crosses denote Bragg peaks of Fe lattice and the modulation wavevectors $Q_{x,y}$ are circled in blue. The Fourier peaks at $\pm Q_x$ are stronger than $\pm Q_y$, indicating the $C_4$ symmetry breaking PDW order. **c,d**, The tunneling spectra taken along $x$ and $y$ directions, respectively ($V_s$ = -100 mV, $I_s$ = 6 nA, $V_{mod}$ = 0.6 mV). The positions of the line profiles are marked by black ($x$ direction) and red ($y$ direction) arrows in **a**. **e**, Measured coherence peak heights in **c** and **d**, along with their Fourier filtered modulation. Black curves represent $x$ direction and red for $y$ direction. The distance is defined relative to the beginning of the arrows. The filtered modulation in $x$ direction has larger amplitude than $y$ direction, supporting the rotational symmetry breaking PDW order. The curves are vertically shifted for comparison.

The interplay between the CDW and induced PDW order is further demonstrated by comparing the modulation amplitude and phase maps of the charge density $\rho$ and the superconducting CPH $H$. The charge density $\rho$-map is obtained by $\rho(\mathbf{r}, V) = I(\mathbf{r}, V) - I(\mathbf{r}, -V) \propto \int_{-eV}^{eV} D(\mathbf{r}, V)$, which is the integrated density of states ($D(\mathbf{r}, V)$) within $\pm eV$. The integrated density of states can eliminate the influence from QPI, which is phase incoherent at different bias voltages[32]. The phenomenological GL theory allows a general analysis on the CDW induced PDW[20]. In the presence of non-zero uniform superconductivity $\Delta_0$ and charge density wave $\rho_{\mathbf{Q}_i}$, the free energy necessarily contains a coupling term amounting to a non-zero PDW component $\Delta_{\mathbf{Q}_i}$, whose amplitude is proportional to $|\Delta_0|$ and $|\rho_{\mathbf{Q}_i}|$. This term also locks the phases of the CDW and induced PDW orders, which leads to a relative phase shift of either 0 or $\pi$, depending on the sign of coupling coefficient (*SI Appendix*). As shown in Fig. S4, the spatial distribution of the CPH modulation amplitude and phase in $x$ direction are almost identical to those of the CDW. The features of the CDW phase, including the phase slip domain walls and $2\pi$ phase winding vortices, are inherited by the PDW phase, further supporting the rotational symmetry breaking (or smectic) PDW is induced by the smectic CDW order.

**Discussion.** As shown in Figs. 2d-f, the anisotropy of the $4a_{Fe}$ CDW is energy dependent, which reaches the maximum near the superconducting gap energy (about 10 meV) and disappears at the energy (about 30 meV) much smaller than the energy where the CDW disappears (> 100 meV), suggesting that the anisotropy of the CDW is strongly correlated to the superconductivity. To further reveal the relationship, we suppress the superconductivity by applying a 14 T magnetic field and measure the evolution of the anisotropy. Figures. 4a-b show the $4a_{Fe}$ charge modulation at 9 mV (near the superconducting energy gap) with and without magnetic field. Figures. 4c-d presents the spatial distribution of the directional order parameter extracted from Figs. 4a-b. The histograms of the directional order parameter under 0 and 14 T are plotted in Fig. 4e, which show histogram peaks at $N^g_{0T} \approx -0.26$ and $N^g_{14T} \approx -0.15$ with and without magnetic field, respectively. The distribution is shifted towards $N^g = 0$ under 14 T, indicating the rotational symmetry breaking is suppressed by the magnetic field. Moreover, defects that locally suppress the superconductivity also weaken the PDW modulation and pin the directionality domain walls where $N^g = 0$ (Fig. S5). Considering that there is no evidence of spin density wave in 1-UC Fe(Te,Se) film, normally the $4a_{Fe}$ CDW order is not expected to be influenced by the magnetic field. We thus propose a heuristic free-energy analysis of the anisotropy of CDW due to the interplay between the CDW and the induced PDW state. The related free energy density is:

$$f = f_{\text{PDW}} + f_{\text{CDW}} + f_c$$

where $f_{\text{PDW}}$ and $f_{\text{CDW}}$ are free energy densities of PDW and CDW with the same form[27] and $f_c$ is the lowest coupling term between the charge and pair density wave:

$$f_{\text{PDW}} = \frac{\alpha_1}{2}\left(\left|\Delta_{\mathbf{Q}_x}\right|^2 + \left|\Delta_{\mathbf{Q}_y}\right|^2\right) + \frac{\beta_1}{4}\left(\left|\Delta_{\mathbf{Q}_x}\right|^2 + \left|\Delta_{\mathbf{Q}_y}\right|^2\right)^2 + \gamma_1\left|\Delta_{\mathbf{Q}_x}\right|^2\left|\Delta_{\mathbf{Q}_y}\right|^2$$

$$f_{\text{CDW}} = \frac{\alpha_2}{2}\left(\left|\rho_{\mathbf{Q}_x}\right|^2 + \left|\rho_{\mathbf{Q}_y}\right|^2\right) + \frac{\beta_2}{4}\left(\left|\rho_{\mathbf{Q}_x}\right|^2 + \left|\rho_{\mathbf{Q}_y}\right|^2\right)^2 + \gamma_2\left|\rho_{\mathbf{Q}_x}\right|^2\left|\rho_{\mathbf{Q}_y}\right|^2$$

$$f_c = \kappa \sum_{j=x,y}\left(\rho_{\mathbf{Q}_j}\Delta_0^*\Delta_{-\mathbf{Q}_j} + \rho_{\mathbf{Q}_j}^*\Delta_0^*\Delta_{\mathbf{Q}_j}\right) + \text{c.c.}$$

where $\alpha_{1,2}$, $\beta_{1,2}$ and $\gamma_{1,2}$ are expansion coefficients. Considering that the ground state in our 1-UC Fe(Te,Se) film is a primary CDW order, the signs of coefficients $\alpha_1$ and $\alpha_2$ are expected to be positive and negative, respectively. Minimizing the free energy $f$, the PDW order parameters $\left|\Delta_{\mathbf{Q}_i}\right|$ ($i = x, y$) are determined by $\frac{4|\kappa||\Delta_0|}{\alpha_1}\left|\rho_{\mathbf{Q}_i}\right|$, i.e., the PDW is induced by the CDW order $\left|\rho_{\mathbf{Q}_i}\right|$ and uniform superconductivity $|\Delta_0|$. The induced PDW has an important effect in that it leads to an important renormalization of the parameter $\gamma_2$, leading to an effective coefficient $\gamma_2^{\text{eff}} \approx \gamma_2 + \gamma_1\left(\frac{4|\kappa||\Delta_0|}{\alpha_1}\right)^4$. $\gamma_2^{\text{eff}}$ determines the CDW symmetries, $\gamma_2^{\text{eff}} > 0$ for stripe order and $\gamma_2^{\text{eff}} < 0$ for checkerboard order. Since near the superconducting gap energy, a strong stripe CDW order is detected, the effective parameter $\gamma_2^{\text{eff}}$ must be positive. The sign of $\gamma_2$ can be determined by the observation of diminishing anisotropy of the CDW order at energies much larger than the superconducting gap energy, as shown in Fig. 2. This implies that when the coupling between the PDW and CDW is physically irrelevant, the CDW order is checkerboard-like, corresponding to $\gamma_2<0$. Thus, in order to produce $\gamma_2^{eff}>0$, the coefficient $\gamma_1$ must satisfy the relation $\gamma_1 \left(\frac{4|\kappa||\Delta_0|}{\alpha_1}\right)^4 > |\gamma_2|$. The positive $\gamma_1$ and negative $\gamma_2$ demonstrate that the anisotropy of the PDW order is not inherited from the CDW order, rather, it is the intrinsic directionality of the PDW order that is the origin of the remarkable evolution in the anisotropy of the CDW. When the superconductivity ($|\Delta_0|$) is weakened by external magnetic fields or defects, $\gamma_2^{eff}$ becomes smaller and the anisotropy of the CDW at low energies is weakened, which is consistent with the experimental results.

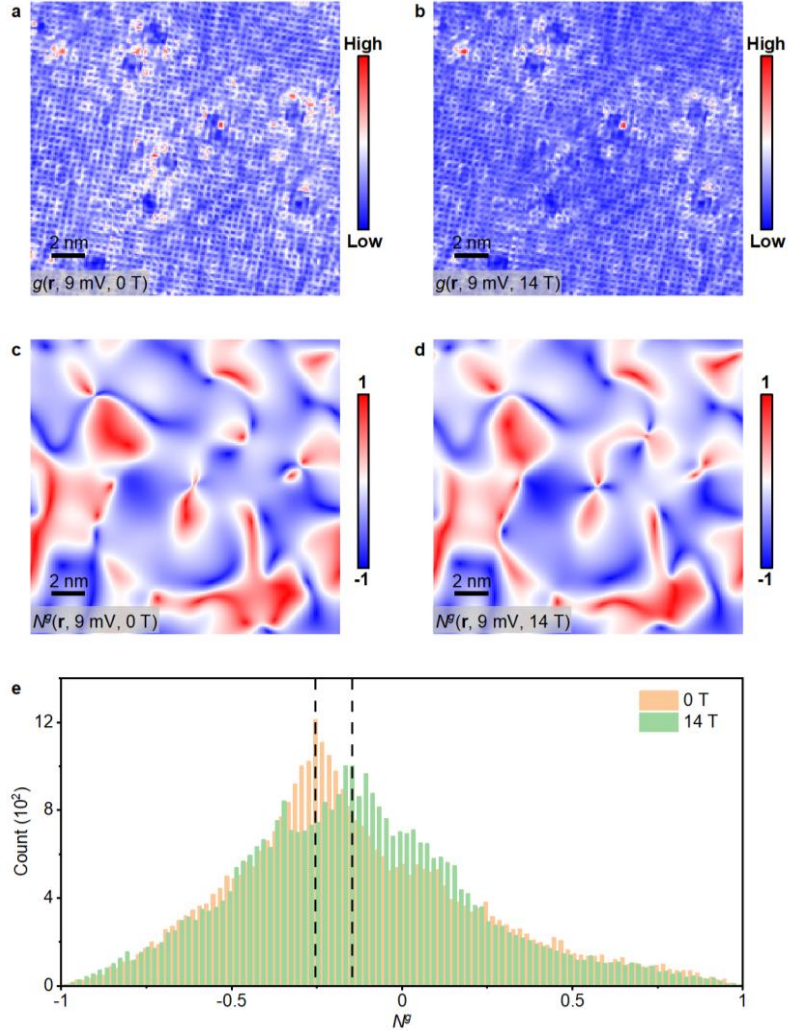

**Fig. 4. Magnetic field suppressed directionality in the CDW order. a,b**, d$I$/d$V$ maps at 9 mV under 0 T (**a**) and 14 T (**b**) measured over a 18.2×17.5 nm$^2$ area ($V_s$ = -100 mV, $I_s$ = 3 nA, $V_{mod}$ = 1 mV). **c,d**, The spatial distribution of the directional order parameter of **a** and **b**. **e**, The statistic histogram of the directional order parameter $N^g$ in **c** (0 T, orange) and **d** (14 T, green). The dashed lines at $N^g_{0T} \approx -0.26$ and $N^g_{14T} \approx -0.15$ denote the peaks of the histograms. The distribution of the $N^g$ is shifted towards $N^g = 0$ under magnetic field, suggesting the CDW directionality is suppressed by magnetic field.

**Conclusion**

In summary, we have observed a commensurate CDW order and CDW-induced PDW order with the same period of $4a_{Fe}$ in the monolayer Fe(Te,Se)/STO. Both of the CDW and PDW orders show rotational symmetry breaking and are electronic smectic orders in analogy to liquid crystals. The energy dependent directionality of the CDW order can be suppressed by magnetic fields and local defects, suggesting a novel mechanism of rotational symmetry breaking due to the coupling between the CDW and PDW order parameters. Our findings provide a new perspective to study the intertwined electronic orders and their interplays in unconventional superconductors.

## Materials and Methods

**Sample preparation and measurement.** The experiments were carried out in an ultra-high vacuum MBE-STM combined system with 15 T magnetic field (Unisoku). The substrate used to grow the sample was Nb-doped SrTiO$_3$(001) (wt 0.7%) with the miscut angle smaller than 0.2°. The substrate was thoroughly degassed at 600°C in ultra-high vacuum MBE chamber with a base pressure about $4 \times 10^{-10}$ mbar. Then the substrate underwent Se-flux treatment at 1000°C for 30 minutes. The 1-UC Fe(Te,Se) film was grown on pretreated substrate by co-evaporating high-purity Fe(99.995%), Te(99.9999%) and Se(99.999%) from Knudsen cells with the substrate at 330°C for 17 minutes. The high-purity Fe, Te, Se were held at 1060°C, 260°C and 120°C, respectively during the depositing process. The Fe(Te,Se) film was annealed at 380°C for 13 hours and 20 minutes and then transported to the *in situ* STM. All STM/S data were acquired at 4.6 K with polycrystalline PtIr tip by the standard lock-in technique. The modulation voltage is $V_{mod}$ = 0.6 mV at 983 Hz unless stated otherwise. WSxM software was used for the post process of STM data[33].

**Measurement of coherence peak height modulation.** We measured tunneling spectrum at each position **r**. For line profiles we took the bias step as 0.2 mV and for the map we took the bias step as 0.5 mV. The steps are smaller than the thermal broadening at 4.6 K ($3.5 k_B T \approx 1.4$ meV). Then we extracted the superconducting coherence peak height (CPH) $H(\mathbf{r})$ at every pixel as follows[10,13]:

1. Take the second derivative of the d$I$/d$V$ spectrum and find the bias $V_0$ that has the minimal value of d$^3I$/d$V^3$.
2. Use three data points in the neighborhood of $V_0$, i.e. {$V_{-1}$, $V_0$, $V_{+1}$}, to fit a quadratic function and take the apex value as the position of coherence peak $V_{CP}$.
3. Extract the d$I$/d$V$ value at $V_{CP}$ by linear interpolation of the measured tunneling spectrum.

Then the CPH map $H(\mathbf{r})$ was further analyzed by Fourier transform and 2D lock-in method.

**Data Availability.** All data are available in the main text or *SI Appendix*.

**ACKNOWLEDGMENTS.** This work was supported by the National Natural Science Foundation of China (Grant No. 11888101), the Innovation Program for Quantum Science and Technology (2021ZD0302403), the China Postdoctoral Science Foundation (2021M700253). Z.W. is supported by the U.S. Department of Energy, Basic Energy Sciences Grant No. DE-FG02-99ER45747.

**References**

1. Fradkin, E., Kivelson, S. A. & Tranquada, J. M. Colloquium: Theory of intertwined orders in high temperature superconductors. *Rev. Mod. Phys.* **87**, 457–482 (2015).
2. Fernandes, R. M., Chubukov, A. V. & Schmalian, J. What drives nematic order in iron-based superconductors? *Nat. Phys.* **10**, 97–104 (2014).
3. Agterberg, D. F. *et al.* The Physics of Pair-Density Waves: Cuprate Superconductors and Beyond. *Annu. Rev. Condens. Matter Phys.* **11**, 231–270 (2020).
4. Hoffman, J. E. *et al.* A Four Unit Cell Periodic Pattern of Quasi-Particle States Surrounding Vortex Cores in $Bi_2Sr_2CaCu_2O_{8+\delta}$. *Science* **295**, 466–469 (2002).
5. Hanaguri, T. *et al.* A 'checkerboard' electronic crystal state in lightly hole-doped $Ca_{2-x}Na_xCuO_2Cl_2$. *Nature* **430**, 1001–1005 (2004).
6. Fernandes, R. M. *et al.* Iron pnictides and chalcogenides: a new paradigm for superconductivity. *Nature* **601**, 35–44 (2022).
7. Böhmer, A. E., Chu, J.-H., Lederer, S. & Yi, M. Nematicity and nematic fluctuations in iron-based superconductors. *Nat. Phys.* **18**, 1412–1419 (2022).
8. Hamidian, M. H. *et al.* Detection of a Cooper-pair density wave in $Bi_2Sr_2CaCu_2O_{8+x}$. *Nature* **532**, 343–347 (2016).
9. Ruan, W. *et al.* Visualization of the periodic modulation of Cooper pairing in a cuprate superconductor. *Nat. Phys.* **14**, 1178–1182 (2018).
10. Edkins, S. D. *et al.* Magnetic field–induced pair density wave state in the cuprate vortex halo. *Science* **364**, 976–980 (2019).
11. Du, Z. *et al.* Imaging the energy gap modulations of the cuprate pair-density-wave state. *Nature* **580**, 65–70 (2020).
12. Wang, S. *et al.* Scattering interference signature of a pair density wave state in the cuprate pseudogap phase. *Nat. Commun.* **12**, 6087 (2021).
13. Liu, Y. *et al.* Discovery of a pair density wave state in a monolayer high-Tc iron-based superconductor. Preprint at https://doi.org/10.48550/arXiv.2209.04592 (2022).
14. Liu, X., Chong, Y. X., Sharma, R. & Davis, J. C. S. Discovery of a Cooper-pair density wave state in a transition-metal dichalcogenide. *Science* **372**, 1447–1452 (2021).
15. Agterberg, D. F. & Tsunetsugu, H. Dislocations and vortices in pair-density-wave superconductors. *Nat. Phys.* **4**, 639–642 (2008).
16. Berg, E., Fradkin, E. & Kivelson, S. A. Pair-Density-Wave Correlations in the Kondo-Heisenberg Model. *Phys. Rev. Lett.* **105**, 146403 (2010).
17. Lee, P. A. Amperean Pairing and the Pseudogap Phase of Cuprate Superconductors. *Phys. Rev. X* **4**, 031017 (2014).
18. Chen, H. *et al.* Roton pair density wave in a strong-coupling kagome superconductor. *Nature* **599**, 222–228 (2021).
19. Gu, Q. *et al.* Detection of a Pair Density Wave State in $UTe_2$. Preprint at https://doi.org/10.48550/arXiv.2209.10859 (2023).
20. Berg, E., Fradkin, E. & Kivelson, S. A. Theory of the striped superconductor. *Phys. Rev. B* **79**, 064515 (2009).
21. Li, F. *et al.* Interface-enhanced high-temperature superconductivity in single-unit-cell $FeTe_{1-x}Se_x$ films on SrTiO3. *Phys. Rev. B* **91**, 220503 (2015).
22. Chen, C. *et al.* Atomic line defects and zero-energy end states in monolayer Fe(Te,Se) high-

temperature superconductors. *Nat. Phys.* **16**, 536–540 (2020).

23. Li, Y. *et al.* Anisotropic Gap Structure and Sign Reversal Symmetry in Monolayer Fe(Se,Te). *Nano Lett.* **23**, 140–147 (2023).
24. Chen, C., Liu, C., Liu, Y. & Wang, J. Bosonic Mode and Impurity-Scattering in Monolayer Fe(Te,Se) High-Temperature Superconductors. *Nano Lett.* **20**, 2056–2061 (2020).
25. Lawler, M. J. *et al.* Intra-unit-cell electronic nematicity of the high-$T_c$ copper-oxide pseudogap states. *Nature* **466**, 347–351 (2010).
26. Li, W. *et al.* Stripes developed at the strong limit of nematicity in FeSe film. *Nat. Phys.* **13**, 957–961 (2017).
27. Robertson, J. A., Kivelson, S. A., Fradkin, E., Fang, A. C. & Kapitulnik, A. Distinguishing patterns of charge order: Stripes or checkerboards. *Phys. Rev. B* **74**, 134507 (2006).
28. Fujita, K. *et al.* Direct phase-sensitive identification of a *d*-form factor density wave in underdoped cuprates. *Proc. Natl. Acad. Sci.* **111**, E3026–E3032 (2014).
29. Chen, W. *et al.* Identification of a nematic pair density wave state in $Bi_2Sr_2CaCu_2O_{8+x}$. *Proc. Natl. Acad. Sci.* **119**, e2206481119 (2022).
30. Cho, D., Bastiaans, K. M., Chatzopoulos, D., Gu, G. D. & Allan, M. P. A strongly inhomogeneous superfluid in an iron-based superconductor. *Nature* **571**, 541–545 (2019).
31. Sulangi, M. A., Atkinson, W. A. & Hirschfeld, P. J. Correlations among STM observables in disordered unconventional superconductors. *Phys. Rev. B* **104**, 144501 (2021).
32. Parker, C. V. *et al.* Fluctuating stripes at the onset of the pseudogap in the high-Tc superconductor $Bi_2Sr_2CaCu_2O_{8+x}$. *Nature* **468**, 677–680 (2010).
33. Horcas, I. *et al.* WSXM: A software for scanning probe microscopy and a tool for nanotechnology. *Rev. Sci. Instrum.* **78**, 013705 (2007).

# Supporting Information for

# Intertwined charge and pair density orders in a monolayer high-$T_c$ iron-based superconductor

Tianheng Wei, Yanzhao Liu, Wei Ren, Ziqiang Wang & Jian Wang

**Corresponding author:** Jian Wang
**Email:** jianwangphysics@pku.edu.cn (J.W.)

**This PDF file includes:**

    Supporting text
    Figures S1 to S7
    SI References

**Phase locking between CDW and induced PDW order**

In the presence of uniform component of superconductivity, the lowest-order coupling free energy density of the CDW and PDW order parameter is given by:

$$f_c = \kappa \sum_{j=x,y} \left( \rho_{\mathbf{Q}_j} \Delta_0^* \Delta_{-\mathbf{Q}_j} + \rho_{\mathbf{Q}_j}^* \Delta_0^* \Delta_{\mathbf{Q}_j} \right) + \text{c.c.} \propto \sum_{j=x,y} \kappa |\rho_{\mathbf{Q}_j} \Delta_0 \Delta_{\mathbf{Q}_j}| \cos\left( \varphi_{\mathbf{Q}_j}^{\text{CDW}} - \frac{\varphi_{\mathbf{Q}_j}^{\text{PDW}} - \varphi_{-\mathbf{Q}_j}^{\text{PDW}}}{2} \right)$$

Experimentally, the phase we extracted from STM data using 2D lock-in method is $\phi_{\mathbf{Q}_j}^{\text{CDW}} = \varphi_{\mathbf{Q}_j}^{\text{CDW}}$ and $\phi_{\mathbf{Q}_j}^{\text{PDW}} = \frac{\varphi_{\mathbf{Q}_j}^{\text{PDW}} - \varphi_{-\mathbf{Q}_j}^{\text{PDW}}}{2}$, which are exactly the two phases appearing in the expression above. To minimize the free energy, $\cos(\phi_{\mathbf{Q}_j}^{\text{CDW}} - \phi_{\mathbf{Q}_j}^{\text{PDW}})$ should be ±1 (depending on the sign of $\kappa$), which locks the phase difference between the CDW and PDW order to be either 0 or $\pi$. In our results, the phase difference is 0 as shown in Fig. S4.

**Possible origins of the CDW order**

The strong bias-dependency of the $4a_{\text{Fe}}$ modulation intensity can exclude the possibility of surface reconstruction, which is expected to be energy independent. And the absence of modulation signal in the topographic image as the bias voltage was set to 500 mV further confirms an electronic feature instead of the surface reconstruction (Fig. S6). Another possible origin is the magnetic order, which is proximity to the superconducting phase in iron-based superconductors. Although the possibility of spin density wave (SDW) has been discussed in FeSe/STO thin films[1], no evidence of SDW was detected in monolayer FeSe/STO or Fe(Te,Se)/STO. Besides, the $4a_{\text{Fe}}$ modulation shows little difference at high magnetic field (14 T) compared to 0 T, making the magnetic order origin less probable. Further experimental investigations using the magnetic tip may be helpful to identify the influence of magnetic order.

Interestingly, the d$I$/d$V$ map at 100 mV $g(\mathbf{r}, 100 \text{ mV})$ (Fig. S7a) shows an obvious dumbbell structure in each $4a_{\text{Fe}}$ modulation supercell. The dumbbell structures have identical orientation along one of the Te/Se lattice directions and are centered at a Fe site, which is clearly shown in the zoom-in map (Fig. S7a inset) and sketched in Fig. S7b. The Fourier transform $g(\mathbf{q}, 100 \text{ mV})$ illustrates 8 extra peaks (red circles) in addition to the $4a_{\text{Fe}}$ modulation wavevectors (blue circles), which are induced by the form factor of the detailed dumbbell structure (Fig. S7c). Similar dumbbell structure has been widely observed in monolayer FeSe/STO topography and interpreted as Fe site defect. Besides, the supercells of commensurate CDW modulation always centered at a Fe site, showing lattice-locked behavior. These phenomena suggest the CDW order may originate from the spontaneous symmetry breaking of the Fe lattice. However, further theoretical and experimental investigations are required to reveal the exactly microscopic mechanism of the observed CDW order in 1-UC Fe(Te,Se)/STO.

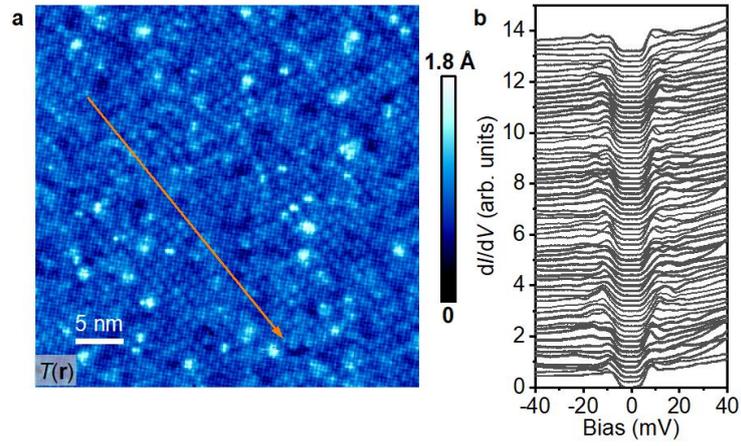

**Fig. S1. Uniform superconductivity of the monolayer Fe(Te,Se)/STO. a**, A typical topographic image of the monolayer Fe(Te,Se)/STO film ($V_s$ = 40 mV, $I_s$ = 100 pA). **b**, Waterfall plot of d$I$/d$V$ spectra measured along the orange arrow in **a** ($V_s$ = 40 mV, $I_s$ = 500 pA, $V_{mod}$ = 0.8 mV). The average of the spectra is the curve shown in Fig. 1**b**.

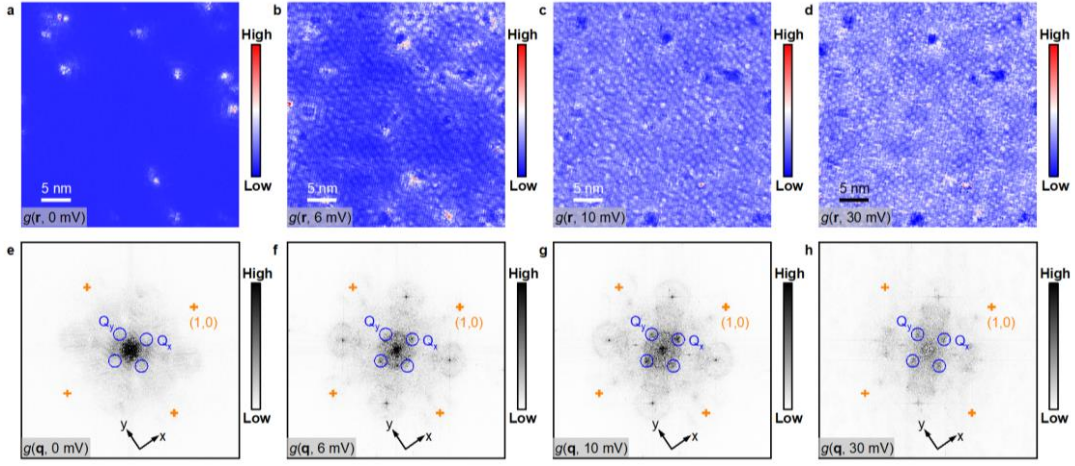

**Fig. S2. Energy dependence of the 4a$_{Fe}$ modulation in LDOS maps. a-d,** d$I$/d$V$ maps measured at Fermi surface (0 mV, **a**), superconducting gap edge (6 mV, **b**), near the coherence peak (10 mV, **c**) and much larger than superconducting gap energy (30 mV, **d**), respectively ($V_s$ = -100 mV, $I_s$ = 2 nA, $V_{mod}$ = 0.6 mV). **e-f,** The magnitude of Fourier transform of **a-d**. Orange crosses denote Bragg peaks of Fe lattice and the modulation wavevectors Q$_{x,y}$ are circled in blue. The modulation is invisible at Fermi energy and emerges when non-zero LDOS is induced by increasing the bias voltage. Then the modulation signal shows the maximum around the superconducting coherence peak and becomes weaker as the bias voltage increases beyond the superconducting gap energy.

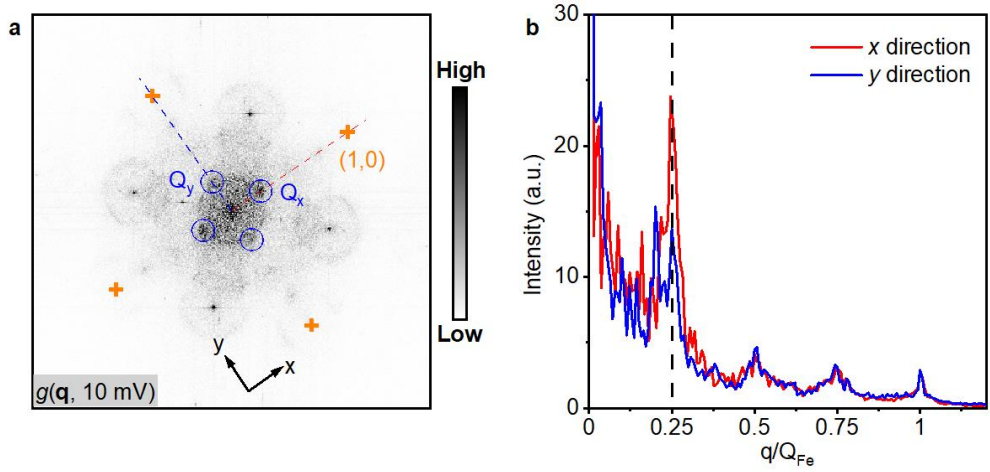

**Fig. S3. Anisotropy of 4a$_{Fe}$ LDOS modulation in *x* and *y* directions. a**, The magnitude of Fourier transform of d*I*/d*V* map at 10 mV shown in the inset of Fig. 1e. The CDW wavevectors are marked by blue circles. **b**, The line profiles of *g*(**q**, 10 mV) along *x* (red dashed line) and *y* (blue dashed line) directions in **a**. The Fourier peak (marked by black dashed line) in *x* direction is stronger than *y* direction.

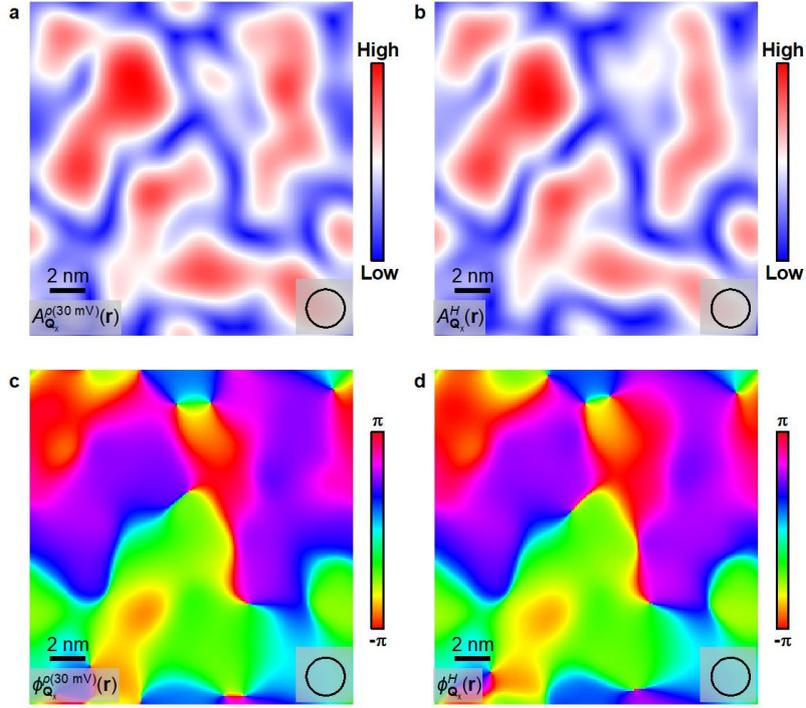

**Fig. S4. Amplitude and phase locking between charge density and superconducting coherence peak height modulation. a,b**, The spatial distribution of the modulation amplitude of $\rho$-map (**a**) and CPH map (**b**) at $\mathbf{Q}_x$ measured at the same area of Fig.3 ($V_s$ = -100 mV, $I_s$ = 6 nA, $V_{mod}$ = 0.6 mV). **c,d**, The spatial distribution of the modulation phase of $\rho$-map (**c**) and CPH map (**d**) at $\mathbf{Q}_x$. Similar amplitudes and locked phases suggest the PDW order is induced by the CDW order. The black circles denote the averaging length scales with their radii equal to the modulation period $4a_{Fe}$.

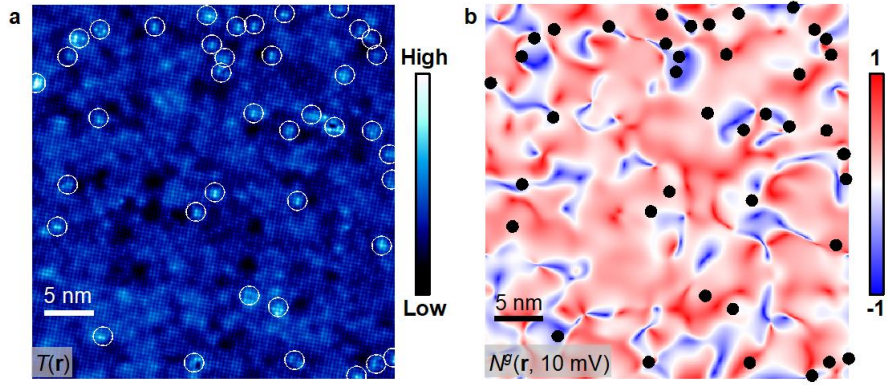

**Fig. S5. The pinning of directionality domain walls by defects. a**, The topographic image measured at the same area in Fig. 1**c** ($V_s$ = -100 mV, $I_s$ = 2 nA). The defects are circled in white. **b**, The spatial distribution of the directional order parameter $N^g(\mathbf{r}, 10\text{ mV})$. The black dots mark the defects positions in **a**. Almost all defects locate near the directionality domain walls (white strings in **b**) where $N^g = 0$, suggesting the directionality domain walls are pinned by defects.

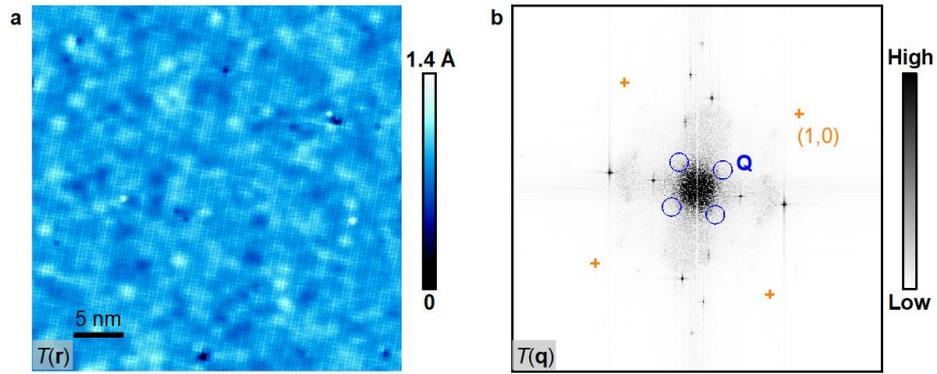

**Fig. S6. Absence of 4a$_{Fe}$ modulation in topography measured at $V_s$ = 500 mV. a**, The topographic image measured at the same area of Fig. 1c. ($V_s$ = 500 mV, $I_s$ = 7 nA). **b**, The magnitude of Fourier transform of **a**, with no Fourier peaks at (±0.25, 0)Q$_{Fe}$ and (0, ±0.25)Q$_{Fe}$. Orange crosses denote the Bragg peaks of Fe lattice.

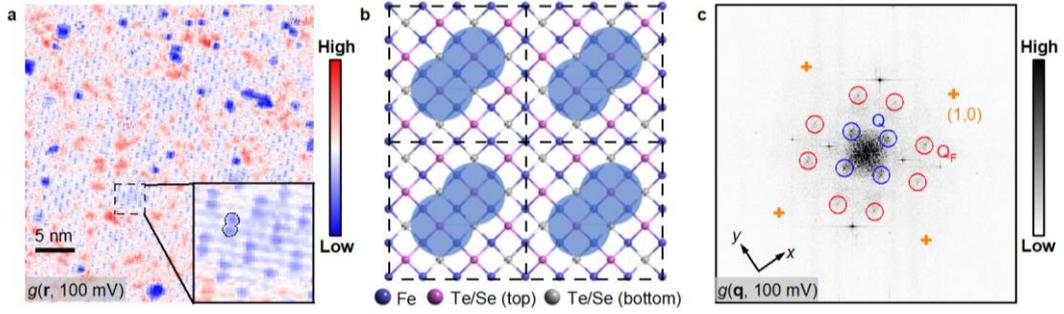

**Fig. S7. Dumbbell structures centered at Fe sites in the supercell. a**, The d$I$/d$V$ map at 100 mV measured in the same area of Fig. 1c ($V_s$ = -100 mV, $I_s$ = 2 nA, $V_{mod}$ = 0.6 mV). Each 4a$_{Fe}$ supercell contains a dumbbell structure centered at Fe site. The inset is the zoom-in clearly showing the dumbbell structures in several 4a$_{Fe}$ supercells. A typical dumbbell structure is marked by black dashed line. **b**, The schematic of dumbbell structures in 4 supercells. **c**, The magnitude of Fourier transform of **a**. Orange crosses denote the Fe lattice Bragg peaks and blue circles denote the CDW wavevector **Q**. In addition, 8 red circles mark the Fourier peaks originating from the form-factor of the dumbbell structure within each supercell.

**Reference**

1. Tan, S. *et al.* Interface-induced superconductivity and strain-dependent spin density waves in FeSe/SrTiO$_3$ thin films. *Nat. Mater.* **12**, 634–640 (2013).